\documentstyle[12pt]{article}
\begin{document}\bibliographystyle{plain}
\begin{titlepage}\renewcommand{\thefootnote}{\fnsymbol{footnote}}
\hfill\begin{tabular}{l}HEPHY-PUB
706/98\\UWThPh-1998-66\\hep-ph/9812526\\November 1998\end{tabular}\\[2.5cm]
\begin{center}
{\Large\bf BOUND STATES BY THE}\\[1ex]{\Large\bf SPINLESS SALPETER EQUATION}\\
\vspace{2cm}{\large\bf Wolfgang LUCHA\footnote[1]{\normalsize\ {\em E-mail\/}:
v2032dac@vm.univie.ac.at}}\\[.5cm]Institut f\"ur Hochenergiephysik,\\
\"Osterreichische Akademie der Wissenschaften,\\Nikolsdorfergasse 18, A-1050
Wien, Austria\\[1.7cm] {\large\bf Franz F.
SCH\"OBERL\footnote[2]{\normalsize\ {\em E-mail\/}:
franz.schoeberl@univie.ac.at}}\\[.5cm]Institut f\"ur Theoretische Physik,\\
Universit\"at Wien,\\ Boltzmanngasse 5, A-1090 Wien,
Austria\\[2cm]\end{center}
\normalsize{\em Invited plenary talk by W. Lucha at the International
Conference ``Nuclear \&~Particle Physics with CEBAF at Jefferson Lab,''
November 3 -- 10, 1998, Dubrovnik, Croatia\/}\begin{center}{\em To appear in
the proceedings\/}\end{center}\vfill\begin{center}{\bf Abstract}\end{center}
\small In quantum theory, bound states are described by eigenvalue equations,
which usually cannot be solved exactly. However, some simple general theorems
allow to derive rigorous statements about the corresponding solutions, that
is, energy levels and wave functions. These theorems are applied to the
prototype of all relativistic wave equations, the spinless Salpeter
equation.\normalsize

\vspace*{6ex}

\noindent{\em PACS\/}: 11.10.St, 03.65.Pm, 03.65.Ge, 12.39.Ki
\renewcommand{\thefootnote}{\arabic{footnote}}\end{titlepage}

\section{INTRODUCTION}In quantum theory, exact statements on solutions of
eigenvalue equations---describing bound states---may be obtained by some
rather simple and elementary methods. Here, these methods are reviewed and
illustrated at the simplest nontrivial (semi-)relativistic wave equation, the
spinless Salpeter equation, which represents a well-defined standard
approximation to the famous Bethe--Salpeter formalism. For more details,
see~Ref.~\cite{Lucha98O}.

\subsection{Relativistic Wave Equation: Spinless Salpeter Equation}

\subsubsection{Bound States in Quantum Field Theory: Bethe--Salpeter
Formalism}The appropriate framework for the description of bound states
within some relativistic quantum field theory is the Bethe--Salpeter
formalism. There a bound state $|{\rm M}\rangle$ (with momentum $K$ and
energy $E$) of, say, fermion and antifermion (with masses $M_1$, $M_2$ and
momenta $P_1$, $P_2$, resp.) is represented by its {\em Bethe--Salpeter
amplitude\/} $\Psi$, defined, in momentum space, in terms of the time-ordered
product of the field operators of~the two bound-state constituents between
vacuum and bound state by $$\Psi(P)=\exp({\rm i}\,K\,X_{\rm
CM})\displaystyle\int{\rm d}^4X\,\exp({\rm i}\,P\,X)\,\langle 0|{\rm
T}(\psi_1(X_1)\,\bar\psi_2(X_2))|{\rm M}(K)\rangle\ ,$$ with the total
momentum $K=P_1+P_2$, relative momentum $P$, center-of-momentum coordinate
$X_{\rm CM}$, and relative coordinate $X\equiv X_1-X_2$ of the bound-state
constituents. The Bethe--Salpeter (BS) amplitude $\Psi$ satisfies the {\em
Bethe--Salpeter equation\/} $$\displaystyle
S_1^{-1}(P_1)\,\Psi(P)\,S_2^{-1}(-P_2)=\frac{{\rm i}}{(2\pi)^4}\int{\rm
d}^4Q\,{\cal K}(P,Q)\,\Psi(Q)\ ,$$ which involves two dynamical
ingredients:\begin{itemize}\item The full fermion propagator $S_i(P)$ of some
particle $i$ ($i=1,2$) of momentum $P$ and mass $M_i$ is usually approximated
by its free form $S_{i,0}^{-1}(P)=-{\rm i}\,(\gamma_\mu\,P^\mu-M_i)$, with
$M_i$ interpreted as some effective (``constituent'') mass and the propagator
understood as an effective one.\item The {\em BS kernel\/} ${\cal K}(P,Q)$ is
defined perturbatively as sum of all two-particle~(BS-) irreducible Feynman
diagrams for two-particle into two-particle scattering (and thus
gauge-dependent!); it is given, e.g., in lowest order QED from one-photon
exchange (in Feynman gauge) by$${\cal
K}(P,Q)=\frac{e^2}{(P-Q)^2}\,\gamma^\mu\otimes\gamma_\mu\ .$$\end{itemize}
The BS equation is formally exact. Nevertheless, it faces some well-known
{\bf problems}:\begin{itemize}\item There is no means to compute the BS
kernel beyond perturbation theory.\item In general, it is not possible to
find the exact solutions of the BS equation (except for very few special
cases, like the famous Wick--Cutkosky model for two scalar particles
interacting by exchange of some massless scalar particle).\item In
non-Abelian gauge theories like QCD, free propagators are incompatible with a
confining BS kernel because Dyson--Schwinger equations connect propagators
and kernel.\end{itemize}\newpage\noindent The usual way out is the {\bf
reduction} of the BS equation by a series of
approximations:\begin{enumerate}\item Eliminate any dependence on timelike
variables:\begin{itemize}\item Adhere to the {\em static approximation\/} to
the BS kernel, i.e., assume that the kernel depends only on the relative {\em
three}-momenta of initial and final state: ${\cal K}(P,Q)={\cal K}({\bf
P},{\bf Q})$. (This is tantamount to ignoring all retardation effects by
assuming instantaneous interactions.)\item Define the {\em equal-time wave
function\/} $\Phi({\bf P})\equiv\displaystyle\int{\rm d}P_0\,\Psi({\bf
P},P_0)$.\end{itemize}This leads to the {\em Salpeter
equation\/}\begin{eqnarray*}\Phi({\bf P})&=&\int\frac{{\rm
d}^3Q}{(2\pi)^3}\left[\frac{\Lambda_1^+\,\gamma_0\,{\cal K}({\bf P},{\bf
Q})\,\Phi({\bf Q})\,\gamma_0\,\Lambda_2^-}{E-\displaystyle\sqrt{{\bf
P}_1^2+M_1^2}-\displaystyle\sqrt{{\bf
P}_2^2+M_2^2}}\right.\\[1ex]&-&\left.\frac{\Lambda_1^-\,\gamma_0\,{\cal
K}({\bf P},{\bf Q})\,\Phi({\bf
Q})\,\gamma_0\,\Lambda_2^+}{E+\displaystyle\sqrt{{\bf
P}_1^2+M_1^2}+\displaystyle\sqrt{{\bf
P}_2^2+M_2^2}}\right],\end{eqnarray*}with the positive/negative-energy
projection operators for particle $i$ ($i=1,2$)
$$\Lambda_i^\pm\equiv\frac{\displaystyle\sqrt{{\bf
P}_i^2+M_i^2}\pm\gamma_0\,(\mbox{\boldmath{$\gamma$}}\cdot{\bf
P}_i+M_i)}{2\,\displaystyle\sqrt{{\bf P}_i^2+M_i^2}}\ .$$ The Salpeter
equation is the equation of motion for the equal-time wave function $\Phi$
with full relativistic kinematics but in static approximation for the kernel
${\cal K}$.\item Neglect the second term on the r.h.s. of the Salpeter
equation by assuming $$E-\sqrt{{\bf P}_1^2+M_1^2}-\sqrt{{\bf P}_2^2+M_2^2}\ll
E+\sqrt{{\bf P}_1^2+M_1^2}+\sqrt{{\bf P}_2^2+M_2^2}\ .$$ This leads to the
{\em reduced Salpeter equation\/} $$\displaystyle\left(E-\sqrt{{\bf
P}_1^2+M_1^2}-\sqrt{{\bf P}_2^2+M_2^2}\right)\Phi({\bf
P})=\displaystyle\int\frac{{\rm
d}^3Q}{(2\pi)^3}\,\Lambda_1^+\,\gamma_0\,{\cal K}({\bf P},{\bf Q})\,\Phi({\bf
Q})\,\gamma_0\,\Lambda_2^-\ .$$\item Neglect all the spin degrees of freedom
of the involved bound-state constituents.\item Assume that the BS kernel
${\cal K}({\bf P},{\bf Q})$ depends only on the difference of the relative
momenta, which means that it is of convolution type: ${\cal K}({\bf P},{\bf
Q})={\cal K}({\bf P}-{\bf Q})$.\item Restrict the whole formalism exclusively
to positive-energy solutions $\psi$.\end{enumerate}Under these simplifying
assumptions and approximations, the BS equation reduces to the {\em spinless
Salpeter equation\/} $$\left[\displaystyle\sqrt{{\bf
P}_1^2+M_1^2}+\displaystyle\sqrt{{\bf P}_2^2+M_2^2}+V({\bf
X})\right]\psi=E\,\psi\ ,$$ involving an interaction potential, $V({\bf X})$,
arising as the Fourier transform of ${\cal K}({\bf P}-{\bf Q})$. In the
center-of-momentum frame of the two bound-state constituents, i.e., for ${\bf
K}={\bf 0}$, this equation reads $H|\psi\rangle=E|\psi\rangle$, with the
Hamiltonian $$H=\sqrt{{\bf P}^2+M_1^2}+\sqrt{{\bf P}^2+M_2^2}+V({\bf X})\ .$$

\subsubsection{Equal-Mass Case}For equal masses of the two bound-state
constituents (i.e., assuming $M_1=M_2=M$), the two-particle spinless Salpeter
equation is equivalent to its one-particle form. To~see this, consider the
two-particle Hamiltonian $H$ with an interaction represented, e.g.,~by a
central potential of power-law form:$$H=2\,\sqrt{{\bf P}^2+M^2}+\sum_{n\in
Z}k_n\,R^n\ ,\quad R\equiv|{\bf X}|\ .$$One may always perform a scale
transformation of the phase-space variables ${\bf X}$, ${\bf P}$ by some
arbitrary scale factor $\lambda$, $${\bf p}=\lambda\,{\bf P}\ ,\quad{\bf
x}=\frac{{\bf X}}{\lambda}\ ,$$ since this rescaling preserves the
fundamental commutation relations: $[{\bf x},{\bf p}]=[{\bf X},{\bf P}]$.
Fix $\lambda$ to $\lambda=2$, which implies $$H=\sqrt{4\,{\bf
P}^2+4\,M^2}+\sum_{n\in Z}k_n\,R^n =\sqrt{{\bf p}^2+4\,M^2}+\sum_{n\in
Z}k_n\,2^n\,r^n\ ,$$identify the one- and two-particle mass and
coupling-strength parameters according~to $m=2\,M$ and $a_n=2^n\,k_n$, $n\in
Z$, and arrive at the equivalent one-particle Hamiltonian $$H=\sqrt{{\bf
p}^2+m^2}+\sum_{n\in Z}a_n\,r^n\ ,\quad r\equiv|{\bf x}|\ .$$

\subsubsection{Prototype: (One-Particle) Spinless Salpeter Equation}In view
of the above, it is sufficient to consider the self-adjoint Hamiltonian
$H=T+V$, with the ``square-root'' operator of the relativistic kinetic energy
of a particle of mass~$m$ and momentum ${\bf p}$, $$T=T({\bf
p})\equiv\sqrt{{\bf p}^2+m^2}\ ,$$ and arbitrary coordinate-dependent, static
interaction-potential operators $V=V({\bf x})$. The {\bf spinless Salpeter
equation} is then nothing else but the eigenvalue equation~for $H$, i.e.,
$H|\chi_k\rangle=E_k|\chi_k\rangle$, $k=0,1,2,\dots$, for Hilbert-space
eigenvectors $|\chi_k\rangle$ and energy eigenvalues
$$E_k\equiv\frac{\langle\chi_k|H|\chi_k\rangle}{\langle\chi_k|\chi_k\rangle}\
.$$It thus represents the simplest relativistic generalization of the
Schr\"odinger equation~of standard nonrelativistic quantum theory. {\bf N.B.}
The semirelativistic Hamiltonian $H$~is a {\em nonlocal\/} operator,
i.e.,\begin{itemize}\item either $T$ {\em in configuration space\/}\item or,
in general, $V$ {\em in momentum space\/}\end{itemize}is nonlocal. Therefore,
it is difficult to obtain rigorous {\em analytic\/} statements on solutions
of this equation of motion.

\subsection{Example: The (Spinless) Relativistic Coulomb Problem}For
illustrative purposes, let us consider a (spherically symmetric) Coulomb
potential, with interaction strength parametrized by some (dimensionless)
coupling constant $\alpha$: $$V({\bf x})=V_{\rm C}(r)=-\frac{\alpha}{r}\
,\quad r\equiv|{\bf x}|\ ,\quad\alpha>0\ .$$For this case, some important
pieces of knowledge have been accumulated until now:\begin{itemize}\item An
examination of the spectral properties of $H$ with a Coulomb potential
reveals \cite{Herbst}\begin{itemize}\item the {\em essential
self-adjointness\/} of $H$ (i.e., its closure is self-adjoint) for
$\alpha\le\displaystyle\frac{1}{2}$;\item the existence of the {\em
Friedrichs extension\/} of $H$ up to a critical value $\alpha_{\rm
c}=\displaystyle\frac{2}{\pi}$;\item a strict {\em lower bound\/} on the
ground-state energy $E_0$ (that is, on $H$), given~by $$\displaystyle E_0\ge
m\,\sqrt{1-\left(\frac{\pi\,\alpha}{2}\right)^2}\quad\mbox{for}\
\alpha<\frac{2}{\pi}\ .$$\end{itemize}In this sense, the Coulombic
Hamiltonian $H$ is a reasonable operator up to~$\alpha_{\rm c}$.\item For
part of the allowed range of $\alpha$, an improved lower bound may be
derived~\cite{Martin}:$$E_0\ge
m\,\sqrt{\frac{1+\sqrt{1-4\,\alpha^2}}{2}}\quad\mbox{for}\
\alpha<\frac{1}{2}\ .$$\item Lower bounds on the spectrum of $H$ may be found
numerically with the help of the {\bf generalized ``local-energy'' theorem}
\cite{Raynal}: Assume that\begin{enumerate}\item the Fourier transform
$\widetilde V({\bf p})$ of $V({\bf x})$ is strictly negative, except
at~infinity, as is the case for an attractive Coulomb potential,\item the
spectrum of $H$ is discrete, and\item the ground state of $H$
exists.\end{enumerate}Define the ``local energy'' $${\cal E}({\bf p})\equiv
T({\bf p})+\frac{\displaystyle\int{\rm d}^3q\,\widetilde V({\bf p}-{\bf
q})\,\phi({\bf q})} {\phi({\bf p})}\ ,$$ with some suitably chosen, positive
trial function $\phi({\bf p})>0$. Then\footnote{\normalsize\ The lower bound
even holds if the spectrum is not purely discrete.}$$\displaystyle\inf_{{\bf
p}}{\cal E}({\bf p})\le E_0<\sup_{{\bf p}}{\cal E}({\bf p})\ .$$This theorem
restricts the ground-state energy $E_0(\alpha)$ as a function of $\alpha$ to
some remarkably narrow band. In particular, at $\alpha_{\rm c}$ one finds
$$0.4825\le\frac{E_0}{m}\le 0.4842910\quad\mbox{for}\ \alpha=\alpha_{\rm c}\
.$$Consequently, $E_0(\alpha=\alpha_{\rm c})$ is definitely
nonvanishing.\end{itemize}However, even for the Coulomb potential the
eigenvalues of $H$ are not known exactly!

\section[]{Analytic Upper Bounds on Energy Levels \cite{Lucha:AUB}}We are
interested in exact analytic expressions for upper bounds on the eigenvalues
of $H$. This only makes sense if the operator under consideration is bounded
from below. Consequently, assume that the potential $V$ is such that $H$ is
bounded from below. For the {\bf Coulomb potential}, this has been rigorously
proven \cite{Herbst}.

\subsection{Bounds on Eigenvalues: Minimum--Maximum Principle}The primary
tool for the derivation of rigorous upper bounds on the eigenvalues~of~a
self-adjoint operator is the well-known {\em minimum--maximum principle\/}.
Among several equivalent formulations the most convenient one for practical
purposes is the following: Let $H$ be a self-adjoint operator bounded from
below with eigenvalues $E_k$, $k=0,1,\dots$, ordered according to $E_0\le
E_1\le E_2\le\dots$, and let $D_d$ be some $d$-dimensional subspace of the
domain of $H$. Then the $k$th eigenvalue $E_k$ (when counting multiplicity)
satisfies$$E_k\le\displaystyle\sup_{|\psi\rangle\in
D_{k+1}}\frac{\langle\psi|H|\psi\rangle}{\langle\psi|\psi\rangle}\quad
\mbox{for}\ k=0,1,2,\dots\ .$$

\subsection{Operator Inequalities}We would like to replace the problematic
kinetic-energy square-root operator in $H$ by some more tractable operator.
One way to achieve this is to use the min--max principle in order to compare
eigenvalues of operators:\begin{itemize}\item Assume the validity of a
(generic) operator inequality of the form $H\le{\cal O}$. Then,
$$E_k\le\sup_{|\psi\rangle\in D_{k+1}}\frac{\langle\psi|H|\psi\rangle}
{\langle\psi|\psi\rangle}\le\sup_{|\psi\rangle\in D_{k+1}}
\frac{\langle\psi|{\cal O}|\psi\rangle}{\langle\psi|\psi\rangle}\ .$$\item
Assume that $D_{k+1}$ is spanned by the first $k+1$ eigenvectors of ${\cal
O}$ (corresponding to the first $k+1$ eigenvalues $\widehat E_0,\widehat
E_1,\dots,\widehat E_k$ of ${\cal O}$ if the latter are ordered according to
$\widehat E_0\le\widehat E_1\le\widehat E_2\le\dots$).
Then$$\sup_{|\psi\rangle\in D_{k+1}}\frac{\langle\psi|{\cal
O}|\psi\rangle}{\langle\psi|\psi\rangle}=\widehat E_k\ .$$\end{itemize}Every
eigenvalue of $H$ is bounded from above by the related eigenvalue of~${\cal
O}$: $E_k\le\widehat E_k$.

\subsection{The ``Schr\"odinger'' Bound}The simplest upper bound on the
eigenvalues of $H$ may be easily found by exploiting the positivity of the
square of the (self-adjoint, since $T$ is self-adjoint) operator
$T-m$:$$0\le(T-m)^2=T^2+m^2-2\,m\,T\equiv{\bf p}^2+2\,m^2-2\,m\,T\
.$$Assuming $m$ to be positive, this may be converted into an operator
inequality for $T$, $T\le m+{\bf p}^2/(2\,m)$, which entails an operator
inequality for $H$: $H\le m+{\bf p}^2/(2\,m)+V$. For the {\bf Coulomb
potential}, the required Schr\"odinger energy eigenvalues are
given~by$$E_{{\rm S},n}=m\left(1-\frac{\alpha^2}{2\,n^2}\right),$$ with the
total quantum number $n$, expressed in terms of the radial quantum number
$n_{\rm r}$ ($n_{\rm r}=0,1,2,\dots$) and the orbital angular momentum $\ell$
($\ell=0,1,2,\dots$) by $n=n_{\rm r}+\ell+1$.

\subsection{A Straightforward Generalization}Real improvement is made by
considering the positivity of the square of the (obviously self-adjoint)
operator $T-\mu$, where $\mu$ is an arbitrary real parameter (with the
dimension of mass):$$0\le(T-\mu)^2=T^2+\mu^2-2\,\mu\,T\equiv{\bf
p}^2+m^2+\mu^2-2\,\mu\,T$$implies a set of operator inequalities for $T$ (see
also Ref.~\cite{Martin}), $T\le({\bf p}^2+m^2+\mu^2)/(2\,\mu)$ $\forall\
\mu>0$. This translates into an operator inequality for $H$: $H\le({\bf
p}^2+m^2+\mu^2)/(2\,\mu)+V$ $\forall\ \mu>0$. Hence, according to the
min--max principle, $E_k$ is bounded from above by the eigenvalue $\widehat
E_{{\rm S},k}(\mu)$ of the Schr\"odinger-like Hamiltonian on the r.h.s. of
this inequality, $E_k\le\widehat E_{{\rm S},k}(\mu)$ $\forall\ \mu>0$, and
thus also by the minimum of these Schr\"odinger-like
bounds:$$E_k\le\displaystyle\min_{\mu>0}\widehat E_{{\rm S},k}(\mu)\ .$$ For
the {\bf Coulomb potential}, the corresponding ``Schr\"odinger'' energy
eigenvalues~are$$\widehat E_{{\rm S},n}(\mu)=\frac{1}{2\,\mu}
\left[m^2+\mu^2\left(1-\frac{\alpha^2}{n^2}\right)\right],$$with the total
quantum number $n=n_{\rm r}+\ell+1$. Minimizing $\widehat E_{{\rm S},n}(\mu)$
w.r.t. $\mu$ then yields $$\displaystyle\min_{\mu>0}\widehat E_{{\rm
S},n}(\mu)=m\,\sqrt{1-\frac{\alpha^2}{n^2}}\quad\forall\ \alpha\le\alpha_{\rm
c}\ .$$These bounds hold for all $\alpha\le\alpha_{\rm c}$ and arbitrary
levels of excitation, and they improve definitely the Schr\"odinger bounds,
for any $n$:$$\min_{\mu>0}\widehat E_{{\rm S},n}(\mu)<E_{{\rm
S},n}\quad\mbox{for}\ \alpha\neq 0\ .$$(For $\mu=m$ one necessarily recovers
the Schr\"odinger case.) The {\bf comparison} of these analytic (!) upper
bounds with their numerically obtained counterparts shows that the relative
error of these analytic bounds (for $\alpha\le0.5$) is for the ground state
($n_{\rm r}=\ell=0$) less than 4.5 $\%$ and for the level $n_{\rm r}=0$,
$\ell=1$ less than 0.1 $\%$.

\subsection{Rayleigh--Ritz Variational Technique}An {\bf immediate
consequence of the min--max principle} is the famous Rayleigh--Ritz
technique: Restrict the operator $H$ to the subspace $D_d$ by orthogonal
projection~$P$ onto $D_d$: $\widehat H:=P\,H\,P$. Let $\widehat E_k$,
$k=0,1,\dots,d-1$, denote all $d$ eigenvalues of $\widehat H$, ordered
according to $\widehat E_0\le\widehat E_1\le\dots\le\widehat E_{d-1}$. The
$k$th eigenvalue (counting multiplicity) of $H$, $E_k$, then satisfies
$E_k\le\widehat E_k$, $k=0,1,\dots,d-1$.

Now, if $D_d$ is spanned by linearly independent basis vectors
$|\psi_k\rangle$, $k=0,1,\dots,d-1$, the eigenvalues $\widehat E$ are simply
determined by diagonalizing the $d\times d$ matrix $(\langle\psi_i|\widehat
H|\psi_j\rangle)$, $i,j=0,1,\dots,d-1$, i.e., as the $d$ roots of the
characteristic equation $$\det\left(\langle\psi_i|\widehat
H|\psi_j\rangle-\widehat E\,\langle\psi_i|\psi_j\rangle\right)=0\ ,\quad
i,j=0,1,\dots,d-1\ .$$

\subsection{Variational Bound for the Ground State}As a first test, let us
apply the Rayleigh--Ritz variational technique to the ground~state, by
considering the case $k=0$ (i.e., an only one-dimensional subspace). In this
case,~the min--max principle reduces to
$$E_0\le\frac{\langle\psi|H|\psi\rangle}{\langle\psi|\psi\rangle}$$(i.e., the
ground-state energy $E_0$ is less than or equal to any expectation value
of~$H$). However, one may consider simultaneously more than one
one-dimensional trial spaces in order to obtain an optimized upper bound
according to the following {\bf prescription}:\begin{enumerate}\item Choose
some suitable set of trial states $\{|\psi_\lambda\rangle\}$ (with elements
distinguished from each other by some variational parameter $\lambda$).\item
Calculate all the expectation values
$E(\lambda)\equiv\langle\psi_\lambda|H|\psi_\lambda\rangle/
\langle\psi_\lambda|\psi_\lambda\rangle$ of $H$ w.r.t. $|\psi_\lambda\rangle$.
\item Determine (from the first derivative of $E(\lambda)$ w.r.t. $\lambda$)
that value $\lambda_{\rm min}$ of $\lambda$ which minimizes
$E(\lambda)$.\item Compute $E(\lambda_{\rm min})$ (i.e., the minimal
expectation value of $H$ in the Hilbert-space subsector of the chosen trial
states).\end{enumerate}The outcome of this procedure is an optimized upper
bound on $E_0$, viz., $E_0\le E(\lambda_{\rm min})$.

In order to get rid of the troublesome square-root operator in $H$, we adopt
a trivial but fundamental inequality for the expectation values of a
self-adjoint operator ${\cal O}$ and its square w.r.t. any state
$|\psi\rangle$ in the domain of ${\cal O}$:
$$\displaystyle\frac{|\langle\psi|{\cal
O}|\psi\rangle|}{\langle\psi|\psi\rangle}\le\sqrt{\frac{\langle\psi|{\cal
O}^2|\psi\rangle}{\langle\psi|\psi\rangle}}\ .$$We consider this inequality,
of course, for $T$, and apply the resulting inequality to
$H$:$$E_0\le\frac{\langle\psi|T+V|\psi\rangle}{\langle\psi|\psi\rangle}
\le\sqrt{\frac{\langle\psi|T^2|\psi\rangle}{\langle\psi|\psi\rangle}}
+\frac{\langle\psi|V|\psi\rangle}{\langle\psi|\psi\rangle}\equiv
\sqrt{\frac{\langle\psi|{\bf p}^2|\psi\rangle}{\langle\psi|\psi\rangle}+m^2}
+\frac{\langle\psi|V|\psi\rangle}{\langle\psi|\psi\rangle}\ .$$ For the {\bf
Coulomb potential}, a special choice for the coordinate-space representation
of trial vectors is suggestive: the well-known (normalized) hydrogen-like
trial functions$$\psi_\lambda({\bf
x})=\sqrt{\displaystyle\frac{\lambda^3}{\pi}}\,\exp(-\lambda\,r)\
,\quad\lambda>0\ .$$This leads to a one-parameter set of upper bounds,
$E_0\le\sqrt{\lambda^2+m^2}-\alpha\,\lambda$ for all $\lambda>0$, with
absolute minimum $E_0\le m\,\sqrt{1-\alpha^2}$, which is identical to the
previous~generalized (operator) bound for $n=1$ and, for $\alpha\neq 0$,
lower and thus better than the Schr\"odinger bound$$E_{{\rm
S},0}=m\left(1-\frac{\alpha^2}{2}\right).$$Thus, the variational technique
yields indeed improved upper bounds on energy levels.

\subsection{Energy Levels at the Critical Coupling Constant of the
Relativistic Coulomb Problem}In order to find, for the {\bf Coulomb
potential}, upper bounds on the energy levels~at~$\alpha_{\rm c}$, we employ
basis vectors $|\psi_k\rangle$ (labelled by a positive integer
$k=0,1,2,\dots$) of $D_d$, given in configuration-space representation by
$$\psi_k(r)=\sqrt{\frac{(2\,m)^{2\,k+2\,\beta+1}}
{4\pi\,\Gamma(2\,k+2\,\beta+1)}}\,r^{k+\beta-1}\exp(-m\,r)\ ,\quad
r\equiv|{\bf x}|\ ,\quad\beta\ge 0\ ,\quad m>0\ .$$

The parameter $\beta$ allows, for a given $\alpha$, of the complete
cancellation of the divergent contributions to the expectation values of the
kinetic energy $T$ for large momenta~$p$~and the Coulomb potential $V_{\rm
C}(r)$ at small distances $r$: $\beta=\beta(\alpha)$. $\beta$ is implicitly
determined as a function of $\alpha$, e.g., for the ground state by
\cite{Durand} $\alpha=\beta\cot(\beta\,\pi/2)$. Consequently,~$\alpha_{\rm
c}$ is approached for $\beta\to 0$. For our $|\psi_k\rangle$, singularities
arise only in the ground-state~matrix elements, i.e., in
$\langle\psi_0|T|\psi_0\rangle$ and $\langle\psi_0|V_{\rm
C}(r)|\psi_0\rangle$. It is a simple task to evaluate all relevant matrix
elements for arbitrary $\beta$.

For simplicity, introduce a dimensionless energy eigenvalue $\varepsilon$ by
$\widehat E=:(2/\pi)\,m\,\varepsilon$. The resulting characteristic equation
for $\beta=0$ is typically of the form $$\det\left(\begin{array}{ccc}4\ln
2-2-\varepsilon&
\displaystyle\frac{\sqrt{2}}{3}-\displaystyle\frac{\varepsilon}{\sqrt{2}}&
\cdots\\[2ex]
\displaystyle\frac{\sqrt{2}}{3}-\displaystyle\frac{\varepsilon}{\sqrt{2}}&
\displaystyle\frac{17}{15}-\varepsilon&\cdots\\[2ex]
\vdots&\vdots&\ddots\end{array}\right)=0\ .$$The roots of this equation
(which may be calculated algebraically up to $d=4$) read,~for $d=1$,
$\varepsilon=2\,(2\ln 2-1)$, entailing the ground-state upper bound $\widehat
E_0/m=0.4918\dots$,~for $d=2$, $$\varepsilon=\frac{1}{15}\left(60\ln
2-23\pm\sqrt{(60\ln 2)^2-4800\ln 2+1649}\right),$$ entailing the ground-state
upper bound $\widehat E_0/m=0.484288\dots$, while, for $d=4$, some lengthy
expression implies the ground-state upper bound $\widehat
E_0/m=0.4842564\dots$. From $d=2$ on, all these analytic bounds lie well
within the numerical ``local-energy range.''

\subsection{Generalized Laguerre Basis}Upper bounds on eigenvalues may be
improved by enlarging $D_d$ to higher dimensions~$d$ or by spanning it by a
more sophisticated set of basis states. A rather popular choice~of trial
functions involves the generalized Laguerre polynomials $L_k^{(\gamma)}(x)$
for parameter $\gamma$: these are orthogonal polynomials defined by the power
series
$$L_k^{(\gamma)}(x)=\sum_{t=0}^k\,(-1)^t\left(\begin{array}{c}k+\gamma\\k-t
\end{array}\right)\frac{x^t}{t!}$$ and normalized, with the weight function
$x^\gamma\exp(-x)$, according to $$\int\limits_0^\infty{\rm
d}x\,x^\gamma\exp(-x)\,L_k^{(\gamma)}(x)\,
L_{k'}^{(\gamma)}(x)=\frac{\Gamma(\gamma+k+1)}{k!}\,\delta_{kk'}\ .$$To
construct a trial vector corresponding to a state with orbital angular
momentum~$\ell$ and its projection $m$, introduce two variational parameters:
$\mu$ (with dimension of mass) and $\beta$ (dimensionless). This state is
defined by the configuration-space representation$$\psi_{k,\ell m}({\bf
x})=\sqrt{\frac{(2\,\mu)^{2\,\ell+2\,\beta+1}\,k!}
{\Gamma(2\,\ell+2\,\beta+k+1)}}\,r^{\ell+\beta-1}\exp(-\mu\,r)\,
L_k^{(2\,\ell+2\,\beta)}(2\,\mu\,r)\,{\cal Y}_{\ell m}(\Omega_{\bf x})\ ,$$
with the orthonormalized spherical harmonics ${\cal Y}_{\ell m}(\Omega)$,
depending on the solid angle~$\Omega$. Normalizability of the trial states
requires $\mu>0$ and $\beta>-1/2$; $\psi_{k,\ell m}({\bf x})$ then satisfies
the (standard) orthonormalization condition $\int{\rm d}^3x\,\psi_{k,\ell
m}^\ast({\bf x)}\,\psi_{k',\ell'm'}({\bf
x)}=\delta_{kk'}\,\delta_{\ell\ell'}\,\delta_{mm'}$.

\subsubsection{Power-Law Potentials}Let us investigate interaction potentials
of power-law form: $V(r)=\sum_na_n\,r^{b_n}$, $r\equiv|{\bf x}|$, where the
real constants $a_n$ and $b_n$ are only constrained by the semiboundedness
of~$H$: $b_n\ge-1$ if $a_n<0$. The matrix elements of $V$ are easily worked
out:\begin{eqnarray*}\langle\psi_i|V|\psi_j\rangle
&=&\sqrt{\frac{i!\,j!}{\Gamma(2\,\ell+2\,\beta+i+1)\,
\Gamma(2\,\ell+2\,\beta+j+1)}}\,\sum_n\,\frac{a_n}{(2\,\mu)^{b_n}}\,
\sum_{t=0}^i\,\sum_{s=0}^j\,\frac{(-1)^{t+s}}{t!\,s!}\\[1ex]
&\times&\left(\begin{array}{c}i+2\,\ell+2\,\beta\\i-t\end{array}\right)
\left(\begin{array}{c}j+2\,\ell+2\,\beta\\j-s\end{array}\right)
\Gamma(2\,\ell+2\,\beta+b_n+t+s+1)\ .\end{eqnarray*}

\subsubsection{Analytically Evaluable Special Cases}\subsubsection*{\bf
Orbital Excitations}Consider just orbital excitations by restricting to
$i=j=0$ but allowing for arbitrary~$\ell$. Only for definiteness, fix $\beta$
to $\beta=1$. In this case, the matrix elements of $T$ are given~by
$$\langle\psi_0|T|\psi_0\rangle=\frac{4^{\ell+2}\,[\Gamma(\ell+2)]^2}
{\sqrt{\pi}\,\Gamma\left(2\,\ell+\frac{7}{2}\right)}\,\mu\,
F\left(-\frac{1}{2},\ell+2;2\,\ell+\frac{7}{2};1-\frac{m^2}{\mu^2}\right),$$
with the hypergeometric series
$$F(u,v;w;z):=\frac{\Gamma(w)}{\Gamma(u)\,\Gamma(v)}\,\sum_{n=0}^\infty\,
\frac{\Gamma(u+n)\,\Gamma(v+n)}{\Gamma(w+n)}\,\frac{z^n}{n!}\ .$$There are
several possibilities to get rid of $F$: In the ultrarelativistic limit
($m=0$),~e.g., the kinetic-energy matrix element
becomes$$\langle\psi_0|T|\psi_0\rangle=\frac{2\,[\Gamma(\ell+2)]^2}
{\Gamma\left(\ell+\frac{3}{2}\right)\Gamma\left(\ell+\frac{5}{2}\right)}\,\mu\
.$$For instance, for the {\bf linear potential} $V(r)=a\,r$, $a>0$,
minimizing $\langle\psi_0|H|\psi_0\rangle$ w.r.t. $\mu$ leads to the minimal
upper bound$$\displaystyle\min_{\mu>0}\langle\psi_0|H|\psi_0\rangle=
2\,\sqrt{2\,a}\,\frac{\Gamma(\ell+2)} {\Gamma\left(\ell+\frac{3}{2}\right)}\
.$$In the limit $\ell\to\infty$, this reduces to
$$\lim_{\ell\to\infty}\left(\min_{\mu>0}\langle\psi_0|H|\psi_0\rangle\right)^2
=8\,a\left(\ell+\frac{5}{4}\right),$$describing linear Regge trajectories,
i.e., $[E(\ell)]^2\propto\ell$, in striking accordance with other findings
based on different considerations \cite{Kang,Lucha91}.

\subsubsection*{Radial Excitations}Consider pure radial excitations by
focusing to states with $\ell=0$. The matrix elements of $T$ then typically
take the form\begin{eqnarray*}
\langle\psi_i|T|\psi_j\rangle&=&\sqrt{\frac{i!\,j!}{\Gamma(2\,\beta+i+1)\,
\Gamma(2\,\beta+j+1)}}\,\frac{4^{\beta+1}}{2\pi}\,\mu\,
\sum_{t=0}^i\,\sum_{s=0}^j\,\frac{(-2)^{t+s}}{t!\,s!}\\[1ex]
&\times&\left(\begin{array}{c}i+2\,\beta\\i-t\end{array}\right)
\left(\begin{array}{c}j+2\,\beta\\j-s\end{array}\right)
\Gamma(\beta+t+1)\,\Gamma(\beta+s+1)\,I_{ts}\ ,\end{eqnarray*}with $I_{ts}$,
for $\mu=m$, and for $2\,\beta$ integer (thus non-negative), i.e.,
$2\,\beta=0,1,2,\dots$,~given by\begin{eqnarray*}I_{ts}&=&\frac{1}{2}
\left[\Gamma\left(\frac{2\,\beta+t+s+|t-s|+1}{2}\right)\right]^{-1}
\sum_{n=0}^{|t-s|}\left(\begin{array}{c}|t-s|\\n\end{array}\right)\\[1ex]
&\times& \Gamma\left(\frac{n+1}{2}\right)
\Gamma\left(\frac{2\,\beta+t+s+|t-s|-n}{2}\right)
\cos\left(\frac{n\,\pi}{2}\right)\\[1ex]
&-&\frac{1}{2}\left[\Gamma\left(2\,\beta+t+s+\frac{3}{2}\right)\right]^{-1}
\sum_{n=0}^{2\,\beta+t+s+2}
\left(\begin{array}{c}2\,\beta+t+s+2\\n\end{array}\right)\\[1ex]
&\times&\Gamma\left(\frac{n+1}{2}\right)
\Gamma\left(2\,\beta+t+s+1-\frac{n}{2}\right)
\cos\left(\frac{n\,\pi}{2}\right).\end{eqnarray*}By this, analytic
expressions for the matrix elements $\langle\psi_i|H|\psi_j\rangle$ of $H$
with a power-law potential may be found. Up to $d=4$, these matrix elements
are, at least in~principle, algebraically accessible. For $d>4$, the energy
matrix $(\langle\psi_i|H|\psi_j\rangle)$ must be diagonalized numerically,
without, however, the need of (time-consuming!) integration~procedures.

\end{document}